\documentclass[aps, prd, twocolumn, amsmath, floats,floatfix, superscriptaddress, nofootinbib,
showpacs]{revtex4}

\usepackage{amssymb}
\usepackage{amsmath}
\usepackage{verbatim}
\usepackage{mathrsfs}
\usepackage{amsfonts}
\usepackage{latexsym}
\usepackage{epsfig}
\usepackage{color}
\usepackage{tikz}
\usepackage{graphicx,subfigure}
\usepackage{units}
\usepackage{envmath}
\usepackage{natbib}
\definecolor{verd}{rgb}{0.13, 0.55, 0.13}

\begin{document}

%%%%%%%%%%%%%%%%%
%%%   TITLE   %%%
%%%%%%%%%%%%%%%%%
\title{Dynamical formation of Proca stars and quasi-stationary solitonic objects}

\author{Fabrizio Di Giovanni} 
\affiliation{Departamento de
  Astronom\'{\i}a y Astrof\'{\i}sica, Universitat de Val\`encia,
  Dr. Moliner 50, 46100, Burjassot (Val\`encia), Spain}

\author{Nicolas Sanchis-Gual}
\affiliation{Departamento de
  Astronom\'{\i}a y Astrof\'{\i}sica, Universitat de Val\`encia,
  Dr. Moliner 50, 46100, Burjassot (Val\`encia), Spain}
  
  \author{Carlos A. R. Herdeiro}
\affiliation{Departamento de F\'{\i}sica da Universidade de Aveiro and CIDMA, Campus de Santiago, 
3810-183 Aveiro, Portugal}

\author{Jos\'e A. Font}
\affiliation{Departamento de
  Astronom\'{\i}a y Astrof\'{\i}sica, Universitat de Val\`encia,
  Dr. Moliner 50, 46100, Burjassot (Val\`encia), Spain}
\affiliation{Observatori Astron\`omic, Universitat de Val\`encia, C/ Catedr\'atico 
  Jos\'e Beltr\'an 2, 46980, Paterna (Val\`encia), Spain}

%%%%%%%%%%%%%%%%
%%%   DATE   %%%
%%%%%%%%%%%%%%%%

\date{\today}

%%%%%%%%%%%%%%%%%%%%
%%%   ABSTRACT   %%%
%%%%%%%%%%%%%%%%%%%%

\begin{abstract}  
We perform fully non-linear numerical simulations within the spherically symmetric Einstein-(complex)Proca system. Starting with  Proca field distributions that obey the Hamiltonian, momentum and Gaussian constraints, we show that the self-gravity of the system induces the formation of compact objects, which, for appropriate initial conditions, asymptotically approach stationary soliton-like solutions known as Proca stars. The excess energy of the system is dissipated by the mechanism of \textit{gravitational cooling} in analogy to what occurs in the dynamical formation of scalar boson stars. We investigate the dependence of this process on the phase difference between the real and imaginary parts of the Proca field, as well as on their relative amplitudes. Within the timescales probed by our numerical simulations the process is qualitatively insensitive to either choice: the phase difference and the amplitude ratio are conserved during the evolution. Thus, whereas a truly stationary object is expected to be approached only in the particular case of equal amplitudes and opposite phases, quasi-stationary compact solitonic objects are, nevertheless, formed in the general case. 
\end{abstract}

%%%%%%%%%%%%%%%%
%%%   PACS   %%%
%%%%%%%%%%%%%%%%


%%%%%%%%%%%%%%%%%%%%%%
%%%   MAKE TITLE   %%%
%%%%%%%%%%%%%%%%%%%%%%

\maketitle

%%%%%%%%%%%%%%%%%%%%%%%%%%%%%%%%%%%%%%%%%%%%%%%%%%
\section{Introduction}\label{sec:introduction}
%%%%%%%%%%%%%%%%%%%%%%%%%%%%%%%%%%%%%%%%%%%%%%%%%%
The dawn of gravitational-wave astronomy~\cite{Abbott:2016blz,Abbott:2016nmj,Abbott:2017vtc,Abbott:2017gyy,TheLIGOScientific:2017qsa,Abbott:2017oio}  opens up an observational window to probe the true nature of astrophysical black hole candidates. These are widely believe to be well described, when near equilibrium, by the Kerr metric~\cite{Kerr:1963ud}. But more exotic theoretical possibilities have been put forward, including horizonless compact objects - see $e.g.$~\cite{Visser:1995cc,Mazur:2001fv,Schunck:2003kk,Mathur:2005zp,Gimon:2007ur,Brito:2015}. Such objects have the theoretical appeal of avoiding conceptual issues related to event horizons and spacetime singularities and could, in some circumstances, mimic the phenomenology of black holes~\cite{Cardoso:2016rao}. For most of the proposals, however, a basic caveat concerns dynamical features, including unknown formation scenarios and unhealthy or unknown stability properties. The notable exception, in this respect, are scalar boson stars~\cite{Schunck:2003kk}, which, at least in spherical symmetry, have been show to be perturbatively stable~\cite{Jetzer:1988vr,Gleiser:1988ih,Lee:1988av} and that can form dynamically~\cite{Seidel:1993zk}. Moreover, their dynamics is sufficiently under control, numerics-wise, so that binaries of these objects haven been studied and some of the corresponding gravitational-wave templates have been obtained~\cite{Liebling:2012fv,Bezares:2017mzk,Palenzuela:2017kcg,Helfer:2018vtq}. Even though boson stars may have dynamical~\cite{Cunha:2017qtt} and phenomenological~\cite{Cunha:2017wao} limitations as black-hole mimickers, at present they typify the best understood model of exotic compact objects. 

Scalar boson stars have vector cousins known as Proca stars~\cite{Brito:2015}. Proca stars are akin to the scalar stars in many respects and it is interesting to ask if they also share their dynamical features and, in particular if Proca stars can also form dynamically. Scalar boson star formation was first discussed by Seidel and Suen~\cite{Seidel:1993zk}, who showed that, in spherical symmetry, a generic complex scalar field configuration collapses to form a boson star. Within spherical symmetry, no gravitational radiation is emitted, and the dissipative mechanism that carries away the excess energy, so that a compact object can form, was called \textit{gravitational cooling}, and takes the form of the ejection of scalar particles. Seidel and Suen~\cite{Seidel:1993zk} also observed that a qualitatively similar phenomenon occurs considering a real, rather than a complex scalar field. The complexity of the scalar field is central in obtaining a truly stationary soliton-like solution of the Einstein-(complex)Klein-Gordon field equations~\cite{Herdeiro:2017fhv}. In the real case, only approximately stationary soliton-like solutions exist, dubbed \textit{oscillatons}~\cite{Seidel:1991zh}. Although these are not truly time independent solutions, since they decay in time by leaking part of their scalar field to infinity, the timescale for this process is extremely long~\cite{Page:2003rd} and hence, for most relevant physics, they are \textit{effectively stable}~\cite{Degollado:2018ypf}. Indeed, the fact they form dynamically in much the same way as the truly stationary boson stars supports this effective stability. 

In this paper we shall show that indeed Proca stars can form dynamically in close parallelism to the way scalar boson stars do. Furthermore, for this dynamical formation, the complexity of the Proca field, which is again central in obtaining truly stationary solutions in this model, is not fundamental. If one takes a real Proca field, or if one takes the real and imaginary parts of the Proca field with a more generic phase difference than that necessary for obtaining Proca stars, the dynamical formation of the compact object follows in a qualitatively similar manner. This hints at the existence of a continuous family of quasi-stationary soliton-like solutions in the Einstein-(complex)Proca system, similar to that discussed in~\cite{Hawley:2002zn} for the scalar case, which are effectively stable. 

This paper is organised as follows. In Section~\ref{sec:formalism} we describe the basic equations of the model we shall consider. Then, in Section~\ref{ID} we discuss the initial data that shall be used in our simulations. The numerical framework is briefly described in Section~\ref{sec:NumImpl} and our numerical results are presented in Section~\ref{sec:num_results}. Conclusions are presented in Section~\ref{sec:conclusions}.

%%%%%%%%%%%%%%%%%%%%%%%%%%%%%%%%%%%%%%%%%%%%%%%%%%
\section{Basic equations}
\label{sec:formalism}
%%%%%%%%%%%%%%%%%%%%%%%%%%%%%%%%%%%%%%%%%%%%%%%%%%

We consider the Einstein-(complex)Proca (EcP) system. The system is described by the action $S=\int d^4x\sqrt{-g}\mathcal{L}$ where the Lagrangian density depends on the Proca potential $\mathcal{A}$, and the field strength $\mathcal{F}=d\mathcal{A}$; the Lagrangian density is given by:
\begin{equation} \label{lagrangian}
\mathcal{L}=\frac{R}{16\pi G}-\frac{1}{4}\mathcal{F}_{\alpha\beta}({\mathcal{F}}^{\alpha\beta})^*-\frac{1}{2}\mu^2\mathcal{A}_{\alpha}({\mathcal{A}}^{\alpha})^*,
\end{equation}
where `*'  denotes complex conjugation, $R$ is the Ricci scalar, $G$ is Newton's constant, and $\mu$ is the mass of the field. We assume that spacetime $\mathcal{M}$ can be foliated by spacelike slices $\Sigma_t$ and we write the spacetime metric in 3+1 form as
\begin{equation}
ds^2=-\alpha^2dt^2+\gamma_{ij}(dx^i+\beta^idt)(dx^j+\beta^jdt),
\end{equation}
where $\alpha$ is the lapse function and $\beta^i$ is the shift vector.
We adopt a conformal decomposition of the spatial metric $\gamma_{ij}$
\begin{equation}
\gamma_{ij}=e^{4\chi}\bar{\gamma}_{ij},
\end{equation}
where $\chi=\ln(\gamma/\bar{\gamma})^{1/12}$ is the conformal factor, and $\gamma\equiv \det\gamma_{ij}$, $\bar{\gamma}\equiv \det\bar{\gamma}_{ij}$. Under the assumption of spherical symmetry, the line element can be written as
\begin{equation}
ds^2=e^{4\chi}\left(a(r,t)dr^2+b(r,t)r^2d\Omega^2\right),
\end{equation}
where $a(r,t)$ and $b(r,t)$ are metric functions.
We introduce a background connection, $\hat{\Gamma}^i_{jk}$, which is taken to be that of flat space in spherical coordinates, and we define
\begin{equation}
\Delta\Gamma^i_{jk}\equiv \bar{\Gamma}^i_{jk}-\hat{\Gamma}^i_{jk}
\end{equation}
which, unlike the two connection themselves, transforms as a tensor field. We define the 
BSSN~\cite{Nakamura87,Shibata:1995we,Baumgarte98} auxiliary variables as
\begin{equation} \label{BSSN_variables}
\Delta^i\equiv \bar{\gamma}^{jk}\Delta\Gamma^i_{jk}.
\end{equation}

We adopt Brown's covariant form \cite{Brown:2009,Alcubierre:2010is} of the BSSN formulation, where the evolved fields are the conformally related 3-dimensional metric, the conformal factor $\chi$, the trace of the extrinsic curvature $K$, the independent component of the traceless part of the conformal extrinsic curvature, $A_{a}\equiv A^{r}_{\,\,r}$,  $A_{b}\equiv A^{\theta}_{\,\,\theta}=A^{\varphi}_{\,\,\varphi}$, and the radial component $\Delta^r$ of the additional BSSN variables \eqref{BSSN_variables} - see~\cite{Sanchis-Gual:2017} for further details about the evolution equations of these fields. Besides the gravitational evolution equations we have two constraint equations, the Hamiltonian constraint and momentum constraint which take the form
\begin{align}
\mathcal{H}\equiv &\ \bar{R}-\left(A_a^2+2A_b^2+\frac{2}{3}K^2\right)-16\pi\rho_e=0, \label{Hamiltonian_constraint} \\
\mathcal{M}\equiv &\ \partial_rA_a-\frac{2}{3}\partial_rK+6A_a\partial_r\chi\nonumber \\
	&\ +(A_a-A_b)\left(\frac{2}{r}+\frac{\partial_rb}{b}\right)-8\pi j_r=0 \label{momentum_constraint},
\end{align}
where $\bar{R}$ is the scalar curvature associated with $\bar{\gamma}_{ij}$, and $\rho_e$ and $j_r$ are matter terms defined in Eqs.~\eqref{rho} and \eqref{j} below. 

The Proca field is split into its scalar potential $\Phi$, 3-vector potential $a_i$, and 3-dimensional electric \textbf{E} and magnetic \textbf{B} fields. Following \cite{knapp2002illustrating}, we introduce the auxiliary variable $\Gamma\equiv D^ia_i$, defined by
\begin{equation} \label{D^ia_i}
D^ia_i=\frac{1}{ae^{4\chi}}\left[ \partial_ra_r-a_r\left(\frac{\partial_ra}{2a}-\frac{\partial_rb}{b}-2\partial_r{\chi}-\frac{2}{r}\right)\right] ,
\end{equation}
where $D^i$ is the 3-dimensional covariant derivative with respect to the physical metric. We are considering a spherically-symmetric system and $\textbf{B}=0$, so that the evolution equations for the complex Proca field and the auxiliary variable are:
\begin{align}
\partial_ta_r&= -\alpha(\gamma_{rr}E^r+\partial_r\Phi)-\Phi\partial_r\alpha+\beta^r\partial_ra_r\nonumber\\&+a_r\partial_r\beta^r, \label{ar_evolution} \\
\partial_tE^r&= \alpha(KE^r+{\mu^2}\gamma^{rr}a_r)+\beta^r\partial_rE^r-E^r\partial_r\beta^r, \\
\partial_t\Phi&= \gamma^{rr}a_r\partial_r\alpha+\alpha(K\Phi-\Gamma)+\beta^r\partial_r\Phi, \\
\partial_t\Gamma& =-E^r\partial_r\alpha-\alpha \mu^2\Phi-\alpha D^iD_i\Phi\nonumber\\
&-2\gamma^{rr}\partial_r\Phi\partial_r\alpha-\Phi D^iD_i\alpha+D^i(\mathcal{L}_{\beta}a_i),\label{Gamma_evolution}
\end{align}
where
\begin{equation}
D^i(\mathcal{L}_{\beta}a_i)=D^i\beta^kD_ka_i+\beta^kD_k\Gamma+D^ia_kD_i\beta^k+a_kD^iD_i\beta^k,
\end{equation}
which in spherical symmetry reduces to:
\begin{equation}
\begin{split}
D^i(\mathcal{L}_{\beta}a_i)&=D^i(\beta^r\partial_ra_i)+D^i(a_r\partial_i\beta^r) \\
			&=D^i\beta^r\partial_ra_i+ \beta^r\partial_r\Gamma+D^ia_r\partial_i\beta^r\nonumber\\
&+a_rD^iD_i\beta^r.
\end{split}
\end{equation}

The matter sources $\rho_e$ and $j_r$, as well as $S_a$, $S_b$ and $S$ appearing in the evolution part of Einstein's equations (not shown here; see~\cite{Sanchis-Gual:2017}) are defined as follows
\begin{align}
\rho_e&= n_an_bT^{ab}, \\
j_r&=-\gamma_{ra}n_bT^{ab}, \\
S_a&= T^r_r, \\
S_b&=T^{\theta}_{\theta}=T^{\phi}_{\phi}, \\
S&=\gamma^{ij}S_{ij}=S_a+2S_b,
\end{align}
where $n_a=(-\alpha,0,0,0)$. The stress-energy tensor of the Proca field reads \cite{Sanchis-Gual:2017}
\begin{equation}
\begin{split}
\label{Tab}
T_{ab}=&-\mathcal{F}_{c(a}\mathcal{F}_{b)}^{c \, *}-\frac{1}{4}g_{ab}\mathcal{F}_{cd}(\mathcal{F}^{cd})^* \\
	&+\mu^2\left[\mathcal{A}_{(a}\mathcal{A}_{b)}^*-\frac{1}{2}g_{ab}\mathcal{A}_c(\mathcal{A}^c)^*\right],
\end{split}
\end{equation}
from which we can compute the source terms of the Einstein equations; they are given by
\begin{align}
&8\pi\rho_e= \gamma_{rr}(E^r)^*E^r+\mu^2\left(\Phi^*\Phi+\gamma^{rr}a_r^*a_r\right) \label{rho}\,,\\
&4\pi(\rho_e+S)= \gamma_{rr}(E^r)^*E^r+2\mu^2\Phi^*\Phi\,, \\
&-\frac{8\pi}{3}(S_a-S_b)=\frac{2}{3}\left( \gamma^{rr}\mu^2a_r^*a_r-\gamma_{rr}(E^r)^*E^r\right)\,, \\
&4\pi j_r=\mu^2 \frac{1}{2}(\Phi^* a_r + \Phi a_r^*). \label{j}
\end{align}
where, for instance, $(E^r)^*E^r=\operatorname{Re}(E^r)^2+\operatorname{Im}(E^r)^2$. Observe that \eqref{Tab} reduces to
\begin{equation}
T_{ab} = T_{ab}^{\text{Re}} + T_{ab}^{\text{Im}}.
\end{equation}

The EcP system yields also the ``Gauss" constraint equation that reads
\begin{equation}\label{Gauss_constraint}
\mathcal{G}\equiv D^iE_i+\mu^2\Phi=0. 
\end{equation}

%%%%%%%%%%
\section{Initial data}
\label{ID}
%%%%%%%%%%
To perform evolutions, it is mandatory to first choose an initial configuration for the Proca field that satisfies all constraint equations. In our work, the Hamiltonian constraint \eqref{Hamiltonian_constraint} is solved numerically, using the method described in~\cite{sanchis2015quasistationary}. Following~\cite{Zilhao:2015},  wherein the  Gauss constraint \eqref{Gauss_constraint} is solved analytically,  for the spherically symmetric case we find the following radial profile for the electric field, $E^r$, and for the scalar potential, $\Phi$:
\begin{align}
E^r&=
	\begin{aligned}[t]
	&-\frac{E^r_0\mu^2\sigma^2}{2r^2}\biggl[\exp{\left(-\frac{r_0^2}{\sigma^2}\right)} -\exp{\left[-\frac{(r-r_0)^2}{\sigma^2}\right]} \\
	&+\sqrt{\pi}\frac{r_0}{\sigma}\left({\rm erf}\left(\frac{r_0}{\sigma}\right)+{\rm erf}\left(\frac{r-r_0}{\sigma}\right)\right)\biggr] \,,
	\end{aligned} \\
\Phi&=\frac{\Phi_{0}}{r} \exp{\frac{(r-r_0)^2}{\sigma^2}},
\end{align}
where $E^r_0,r_0,\sigma$, and $\Phi_0$ are constants and `${\rm erf}$' denotes the error function. From now on we will label as $R(r)$ and $R'(r)$ the radial profile of $E^r$ and $\Phi$, respectively.

We take our initial geometry parameters to be:
\begin{equation}\label{parameters}
\begin{split}
\beta^r&=0,  \\
K&=0, \\
a(r,t)&=b(r,t)=1, \\
A_a&=A_b=0, \\
\Delta^r&=0, \\
\alpha&=1. \\
\end{split}
\end{equation}
Then, at time $t=0$ the Proca evolution equations can be written as
\begin{align}
\partial_ta_r&= -(\gamma_{rr}E^r+\partial_r\Phi), \label{ar_equation}\\
\partial_tE^r&= {\mu^2}\gamma^{rr}a_r,  \label{Er_equation}\\ 
\partial_t\Phi&= \Gamma, \label{Phi_equation}\\
\partial_t\Gamma&=- \mu^2\Phi- D^iD_i\Phi. \label{Gamma_equation}	
\end{align}
From~\eqref{Er_equation} we observe that $a_r$ is the time-derivative of $E^r$, multiplied by $\gamma^{rr}\mu^2$. Then, in order to set a phase difference $\delta$ between the real and the imaginary part of the Proca field, we choose the following ansatz for the quantities describing this field:
\begin{align}   \label{first_ansatz}
\operatorname{Re}(E^r)&=A_1 \cos(\omega t+\delta) R(r)\,, \\
\operatorname{Im}(E^r)&= A_2\cos(\omega t+2 \delta) R(r)\,, \\
\operatorname{Re}(a_r)&=-A_1\omega \sin(\omega t+\delta) R(r)\,, \\
\operatorname{Im}(a_r)&=-A_2\omega \sin(\omega t + 2\delta) R(r)\,, \\
\operatorname{Re}(\Phi)&=A_1\cos(\omega t + \delta) R'(r)\,, \\
\operatorname{Im}(\Phi)&=A_2\cos(\omega t +2 \delta) R'(r)\,.
\end{align}
This choice accommodates Eq.~\eqref{Er_equation} if we identify $\mu^2=\omega$. Note that $A_1$ and $A_2$ are the initial amplitudes for the real and imaginary parts, respectively.

With this ansatz, we solve numerically the Hamiltonian constraint, which yields the conformal factor $\chi$ and as a consequence $\gamma^{rr}$; the latter is then used to update the value of  $a_r$, via Eq.~\eqref{Er_equation}.  
%
%by $\gamma^{rr}$. \ch{The last sentence is confusing.} \fdg{Yes it's confusing, maybe we can just don't write that, it's just a detail. What I mean is that from equation \eqref{Er_equation} we have that $a_r$ is not exactly the derivative of $E^r$ but we should also divide it by $\gamma^{rr}$. But $\gamma^{rr}$ is not set from the beginning but we have it only after solving numerically the Hamiltonian constraint. So what I do is to set $a_r$ as the derivative of $E^r$ so I have the initial configuration and I can solve the Hamiltonian constraint, and then when I have $\gamma^{rr}$ I divide $a_r$ by it.}
%
Taking into account the initial parameters as written in Eq. \eqref{parameters}, the momentum constraint \eqref{momentum_constraint} at $t=0$ reduces to $\mathcal{M}\equiv-8\pi j_r=0$ and thus $\Phi^* a_r + \Phi a_r^*=0$ by \eqref{j}. For the case $A_1=A_2$, this constraint can be solved at $t=0$ by requiring $\cos\delta\sin\delta+\cos(2\delta)\sin(2\delta)=0$, which has solutions $\delta=\{0,{\pi}/{2},{\pi}/{3},\pi \}$. Thus, with this setup for a complex Proca field we shall focus on these four cases. We further remark that the solution $\delta=\pi/2$ holds even if $A_1\neq A_2$. 

%%%%%%%%%%%%%%%%%%%%%%%%%%%%%%%%%%%%
\section{Numerical framework}
\label{sec:NumImpl}
%%%%%%%%%%%%%%%%%%%%%%%%%%%%%%%%%%%%

The numerical evolutions of the EcP system discussed in Section~\ref{sec:formalism} are performed using the code in spherical polar coordinates described in~\cite{Montero:2012yr}. For the current simulations, the original code had to be upgraded to account for the Proca field. Details about this upgrade can be found in~\cite{Sanchis-Gual:2017}. The time update of the system of evolution equations (Einstein and Proca) is done using the same type of techniques used in previous works of our group - see, in particular~\cite{Montero:2012yr,sanchis2015quasistationary2, sanchis2015quasistationary}, for complete details. Briefly, we simply point out that the evolution equations are integrated using the second-order Partially Implicit Runge-Kutta method developed in \cite{Isabel:2012arx, Casas:2014}. This method allows to handle singular terms that appear in the evolution equations due to our choice of curvilinear coordinates.  For the simulations we set $\mu=G=c=1$, by using a scaled radial coordinate $r\rightarrow r\mu$ (together with $E\rightarrow E\mu$, $t\rightarrow t\mu$, $\omega\rightarrow \omega/\mu$ where $E$ is the total Proca energy defined by \eqref{Proca_energy}).

The computational domain is covered by an isotropic grid composed of two patches, a geometrical progression in the 
interior part up to a given radius and a hyperbolic cosine outside. On the one hand, using the 
inner grid alone would require too many grid points to place the outer boundary 
sufficiently far from the origin, and hence prevent the effects of possible 
spurious reflections. On the other hand, using only the hyperbolic cosine patch would 
produce very small grid spacings in the inner region of the domain, leading to 
prohibitively small timesteps due to the Courant-Friedrichs-Lewy  condition. Details about the computational grid can be found in~\cite{sanchis2015quasistationary}. In our work the minimum resolution $\Delta r$ 
we choose for the isotropic logarithmic grid is $\Delta r=0.05$. With this 
choice the inner boundary is then set to $r_{\text{min}} = 0.025$ and the outer 
boundary is placed at $r_{\text{max}} = 4180.475$. We choose this value for the outer boundary (the maximum radius of our finite-size  computational grid) so that the light-crossing time of numerical errors propagating from the origin, being reflected by the outer boundary and coming back towards the origin is of the order of the final time of our simulations, $t\sim 8000$. In this way, the region of study is not affected by these errors during the simulation. Our available computational power limits the final time, the resolution and the outer boundary of our simulations. The time step is set to $\Delta t = 0.5 \Delta r$ in order to obtain long-term stable simulations. 

%%%%%%%%%%%%%%%%%%%%%%%%%%%%%%%%%%%%%%%%%%%%%%%%%%%
\section{Results} \label{sec:num_results}
%%%%%%%%%%%%%%%%%%%%%%%%%%%%%%%%%%%%%%%%%%%%%%%%%%

The numerical simulations reported herein start with a spherical cloud of the Proca field, as described in Section~\ref{ID}, which collapses under its own gravity. If the system can dissipate enough energy, via gravitational cooling, a compact object will form: a Proca star~\cite{Brito:2015} or a vector oscillaton. Proca stars have been constructed as stationary solutions of the EcP system using a field ansatz  given in term of two real functions ($A_t$, $A_r$), depending solely on the radial coordinate~\cite{Brito:2015}:
\begin{equation}
\label{1-form}
\mathcal{A}=e^{-i\omega t}\left(iA_tdt+A_rdr\right).
\end{equation}
Illustrative examples of these radial functions, for two spherical Proca star solutions, can be found in Fig.~2 of~\cite{Brito:2015} (but in a different coordinate system). The translation between the two functions ($A_t$, $A_r$) and the value for the Proca field variables described in Eqs.~\eqref{ar_evolution}-\eqref{Gamma_evolution} is provided by
\begin{align}
\Phi&= -n^{\mu}\mathcal{A_{\mu}}, \\
a_i&= \gamma^\mu_i\mathcal{A_{\mu}},\\
E^i&= -i\frac{\gamma^{ij}}{\alpha}\left(D_i(\alpha\Phi)+\partial_ta_j\right),
\end{align}
where $n^{\mu}=\frac{1}{\alpha}(1,-\beta^i)$. Then from \eqref{1-form} we obtain
\begin{align}
\Phi&= -i\frac{A_t}{\alpha}+\frac{\beta^rA_r}{\alpha},\\
a_r&= A_r.
\end{align}
For Proca stars in their ground state, $A_t(r)$ has one node and $A_r(r)$ is nodeless \cite{Brito:2015}. These are the fundamental stable solutions. There are also excited states with more nodes for these functions, but as for scalar boson stars they should be unstable towards decay to the fundamental state. 

\begin{table}[t]
\caption{Spherically symmetric Proca star models. Columns indicate the phase difference between the real and the imaginary parts of the field, $\delta$, the corresponding ratio of their amplitudes, the oscillation frequencies, $\omega_n$, and the initial Proca energy $E_0$.}
\centering 
\begin{tabular}{c c c c c c c c}
\hline
\hline                  
Model & $\delta$ & $A_2/A_1$ & $\omega_1/\mu$ & $\omega_2/\mu$ & $\omega_3/\mu$ & $\omega_4/\mu$ & $E_0\mu$ \\ [0.5ex]
\hline
1 & 0 & 1 &  0.971 & 0.974 & 0.977 & 0.9953 &1.306  \\
2 & $\frac{\pi}{3}$ & 1 & 0.974 & 0.977 & 0.995& - & 1.246   \\
3 & $\frac{\pi}{2}$ & 1 & 0.974 & 0.977 & 0.995 & - & 1.273  \\
4 & $\pi$ & 1 & 0.971 & 0.974& 0.977 &0.9950  & 1.306 \\
5 & Real & 1  & 0.971 & 0.974 & 0.977 & 0.9948 & 1.235 \\
6 & $\frac{\pi}{2}$ & 0.5 & 0.968 & 0.973 & 0.976 & 0.986 & 1.407 \\
7 & $\frac{\pi}{2}$ & 0.9 & 0.974 & 0.976 & 0.9948 & - & 1.319 \\ [1ex]
\hline
\hline
\end{tabular}
\label{table:models}
\end{table}

\begin{figure}[t]
\begin{minipage}{1\linewidth}
\includegraphics[width=0.9\textwidth]{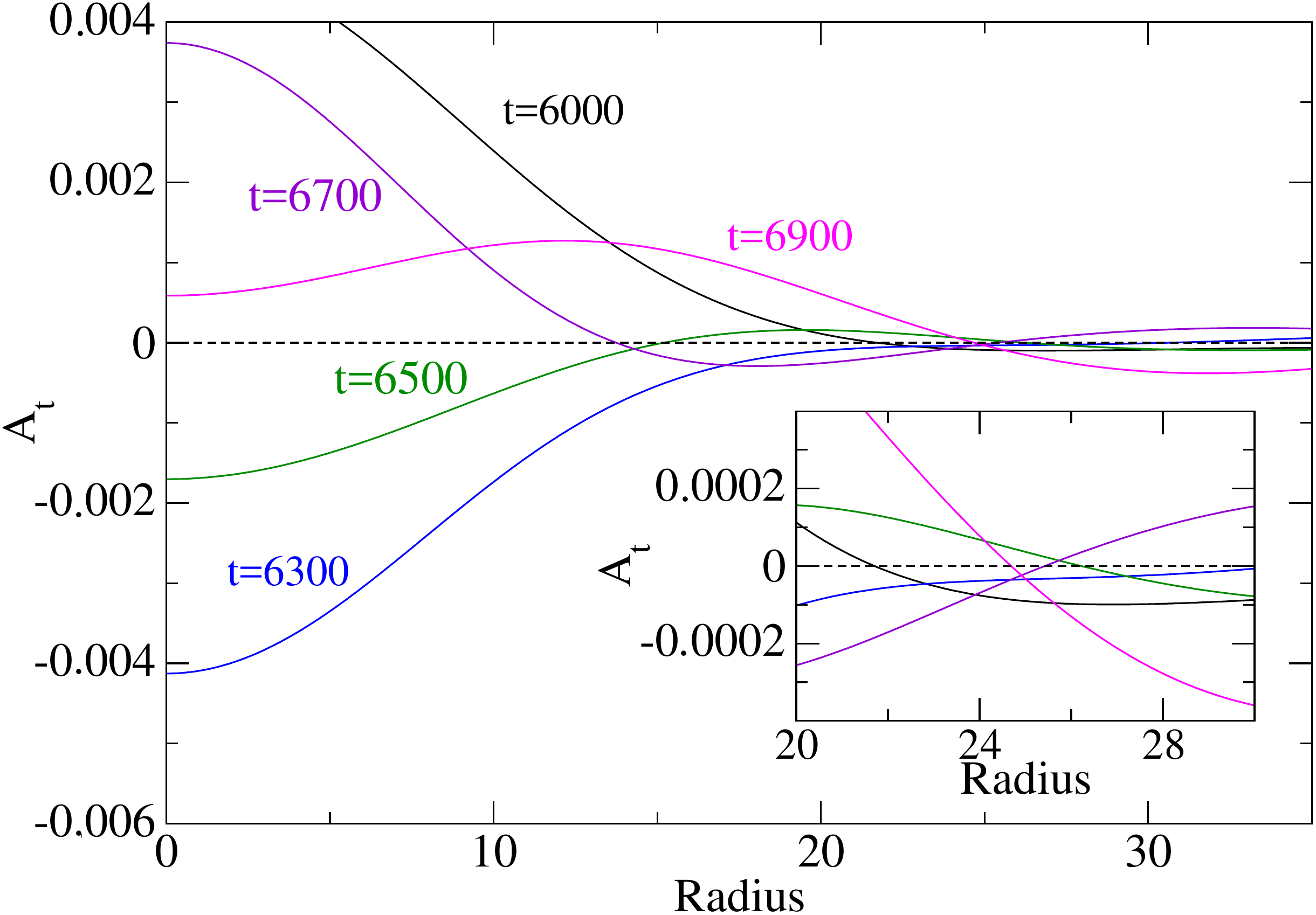} 
\includegraphics[width=0.9\textwidth]{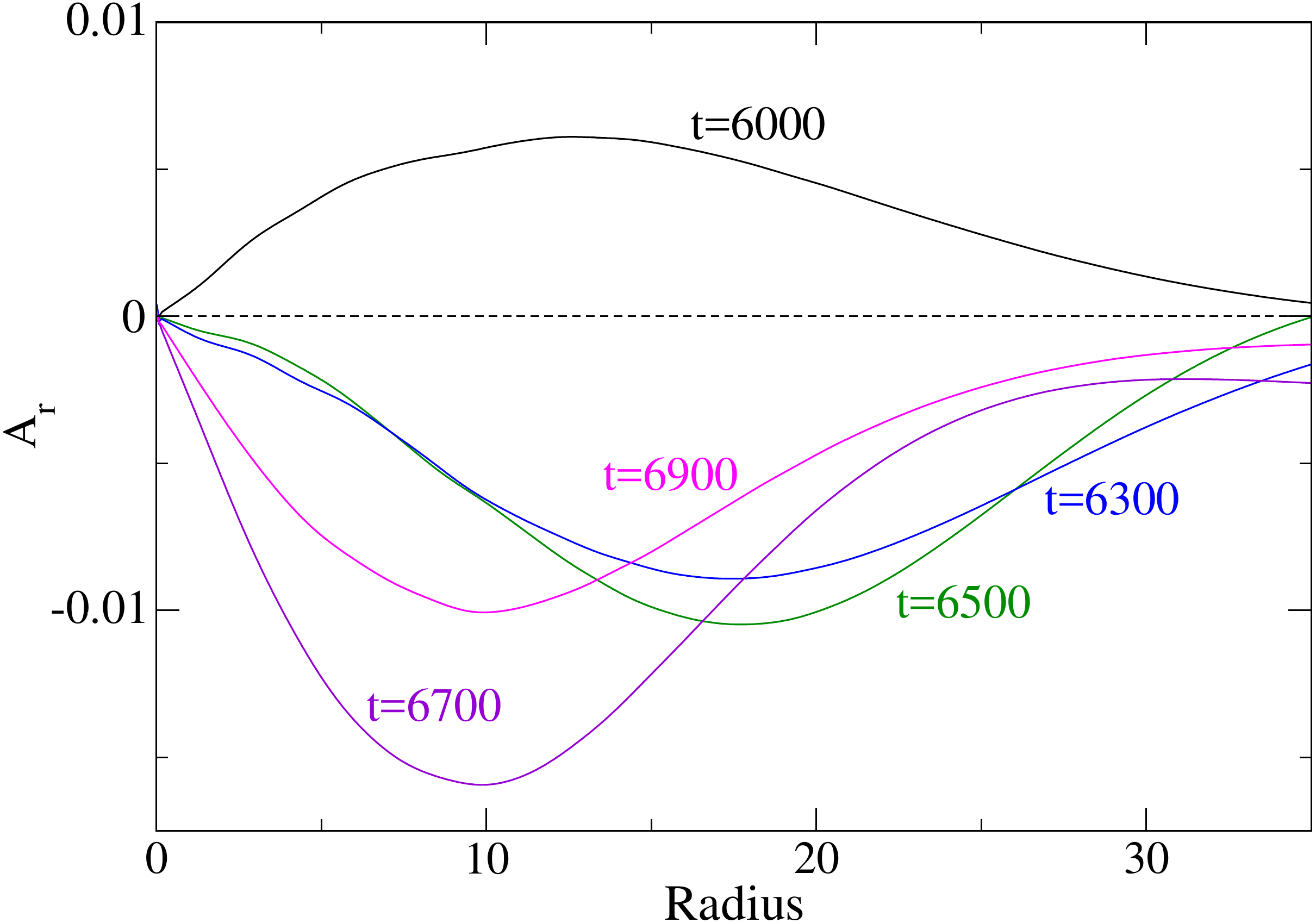} 
\caption{Evolution of the radial profile of $A_t$ (\textit{top panel}) and $A_r$ (\textit{bottom panel}) at intermediate-late times, for model 3. In the inset we focus near the node of $A_t$.}
\label{fig:At-Ar}
\end{minipage}
\end{figure}

We perform evolutions with seven different models, varying the phase difference $\delta$ and the amplitude ratio $A_2/A_1$. All these models have a comparable initial Proca energy calculated as 
\begin{equation} \label{Proca_energy}
E_\infty=\int_{0}^{\infty}\rho_{e}dV,
\end{equation}
and the same parameters for the Gaussian profile, $\sigma=90$ and $r_0=0$. In order to analyze the results of the simulations we extract a time series for the scalar potential $\Phi$ at radius $r=5$. Thus, to identify the frequencies at which the field oscillates we perform a Fast Fourier transform and obtain the power spectrum. We identify for all the models several frequencies of oscillation, labelled $\omega_n$, with $n\in\mathbb{N}$. Table I summarizes the main features of the models. Note that model 5 differs from the rest as it corresponds to a real oscillaton. It is included in our sample for completeness and comparison purposes. In the following, we illustrate our results using models with $\delta={\pi}/{2}$. The behaviour of the other cases is similar, except for the phase difference (cf.~Fig.~\ref{fig:phase} below).

\begin{figure}[t]
\begin{minipage}{1\linewidth}
\includegraphics[width=0.9\textwidth]{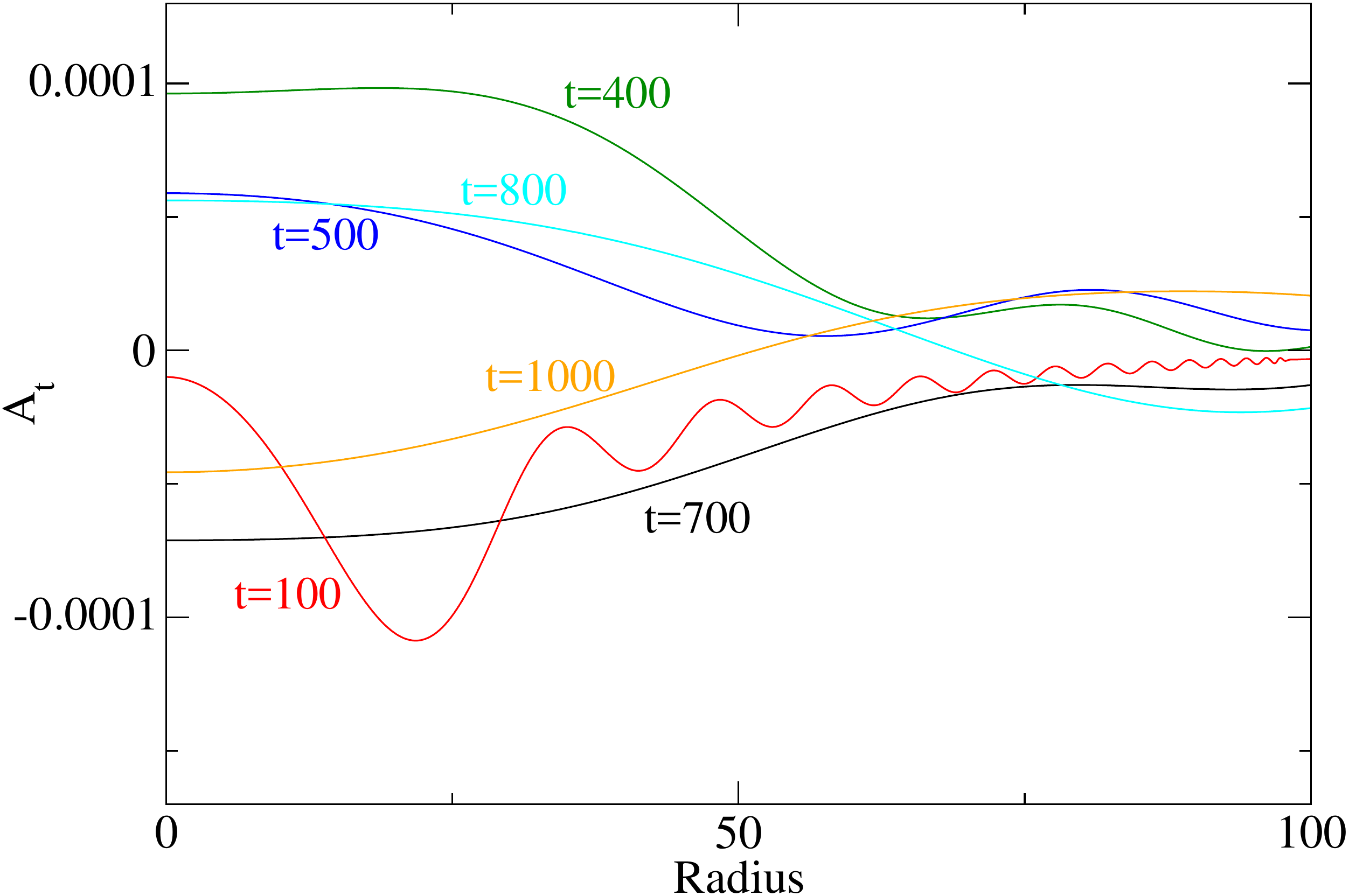} 
\caption{Early time evolution, $t\in[100,1000]$, of the radial profile of $A_t$ for model 3.}
\label{fig:At-beginning}
\end{minipage}
\end{figure}

\begin{figure}[t]
\begin{minipage}{1.0\linewidth}
\includegraphics[width=0.86\textwidth]{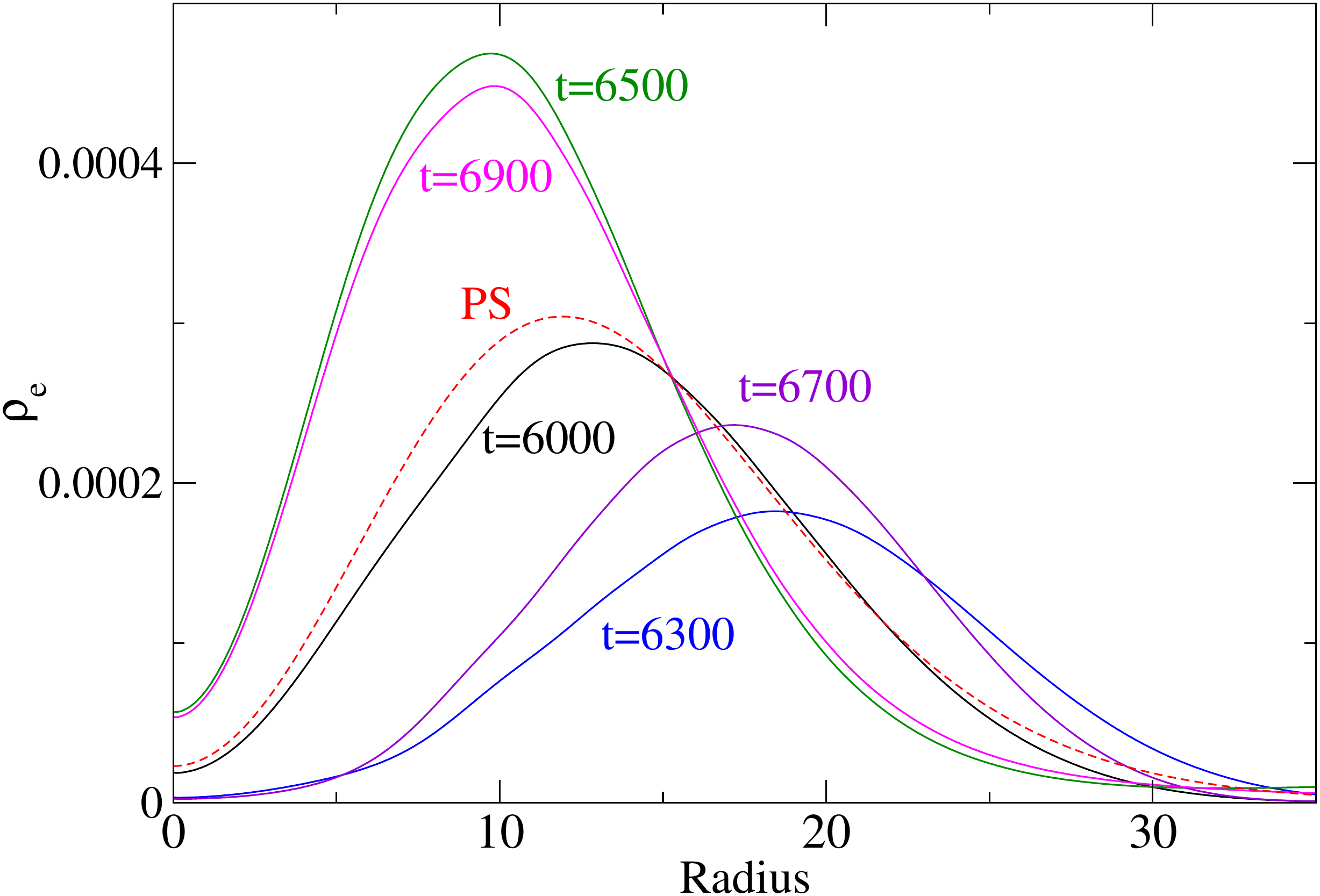} 
\caption{Radial profile of the energy density $\rho_e$ for model 3 at selected times  in between times 6000 and 6900 and for $r\in[0,35]$. During the evolution we can see how the peak amplitude changes in an oscillatory manner, confined in the inner region defined by $r\simeq 30$. The red dashed line corresponds to the radial profile of $\rho_e$ for an isolated, stable Proca star with a similar mass.}
\label{fig:Density2}
\end{minipage}
\end{figure}

In Fig. \ref{fig:At-Ar} we exhibit the evolution of the radial profile of $A_t$ and $A_r$ for model 3 for intermediate-late times. In particular, from the inset in the top panel, a node in $A_t$ can be observed, as expected for a Proca star. In Fig.~\ref{fig:At-beginning} we show the early-time evolution of $A_t$ in between time 100 and 1000. In particular, we note how the radial node is still not present before the formation of the compact object. Fig.  \ref{fig:At-Ar}  suggests a compact object is forming in the volume up to $r\simeq 30$. This is confirmed by the radial profile of the energy density depicted for late times, from $t\sim 6000$,  in 
Fig.~\ref{fig:Density2}.

\begin{figure}
\begin{minipage}{1\linewidth}
\includegraphics[width=0.86\textwidth]{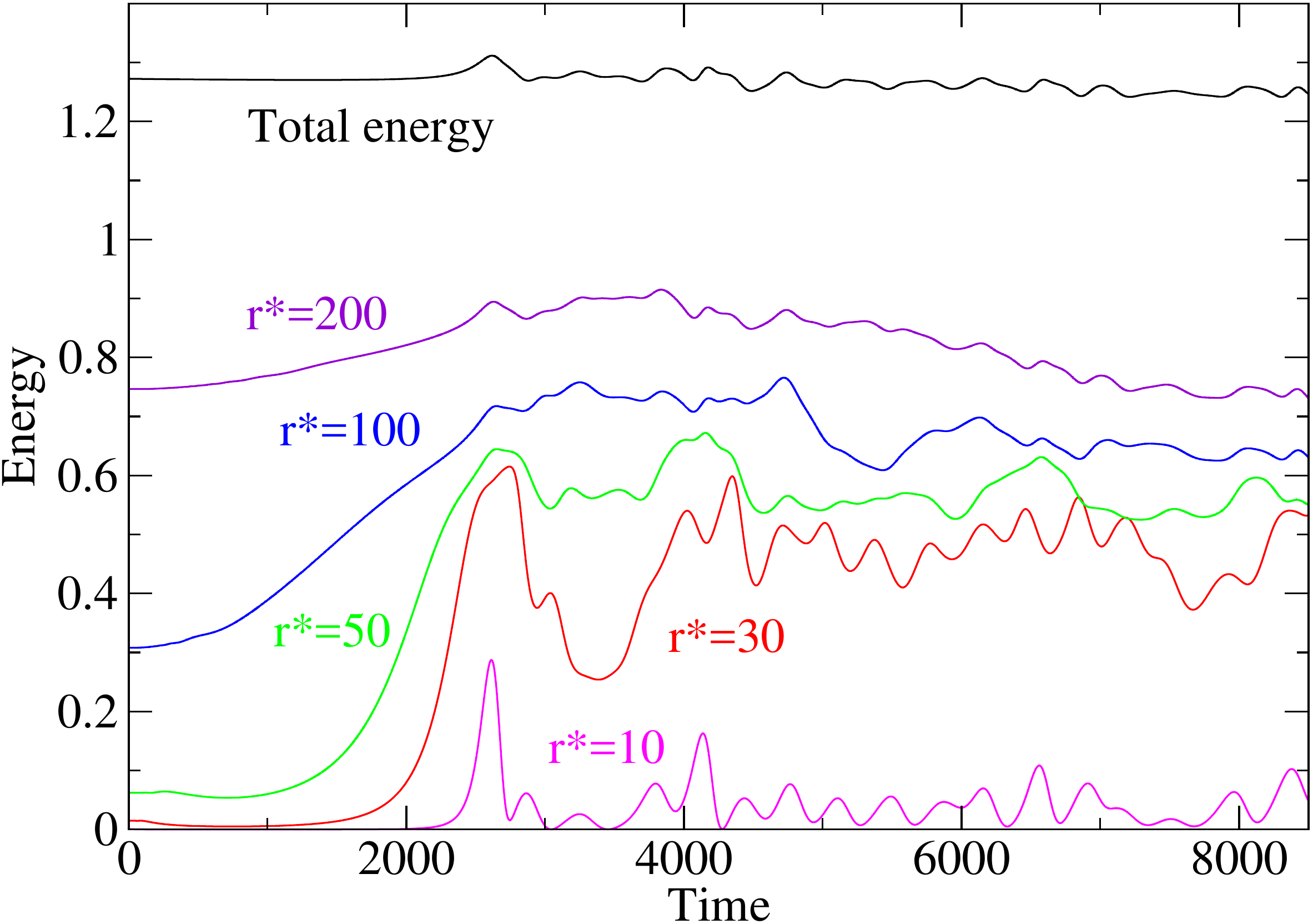} 
\caption{ Proca energy in spheres of different radii, for model 3. The magenta, red, green, blue and violet lines, correspond to, respectively, $r^*=10,30,50,100,200$. The black line represents the total energy.}
\label{fig:Energy}
\end{minipage}
\end{figure}

In Fig. \ref{fig:Energy} we show the time evolution of the total Proca energy contained in spheres of different radii $r^*$, calculated as
\begin{equation} \label{Proca_energy_2}
E_{r^* }=\int_{0}^{r^*}\rho_{e}dV.
\end{equation}
 If we compare the $E_{200}$, $E_{100}$, $E_{50}$ curves to the $E_{30}$ line, we can see that during the evolution, part of the energy escapes from the larger volumes and the four curves seem to slowly converge. The remnant energy is confined to the volume delimited by $r=30$, suggesting the formation of a compact object with a total energy $E_{30}\simeq 0.6$. This is in good agreement with our computed oscillation frequencies (see Table \ref{table:models}) and the mass-frequency relation shown in Fig.~1 of \cite{Brito:2015}. 

\begin{figure}
\begin{minipage}{1\linewidth}
\includegraphics[width=0.9\textwidth]{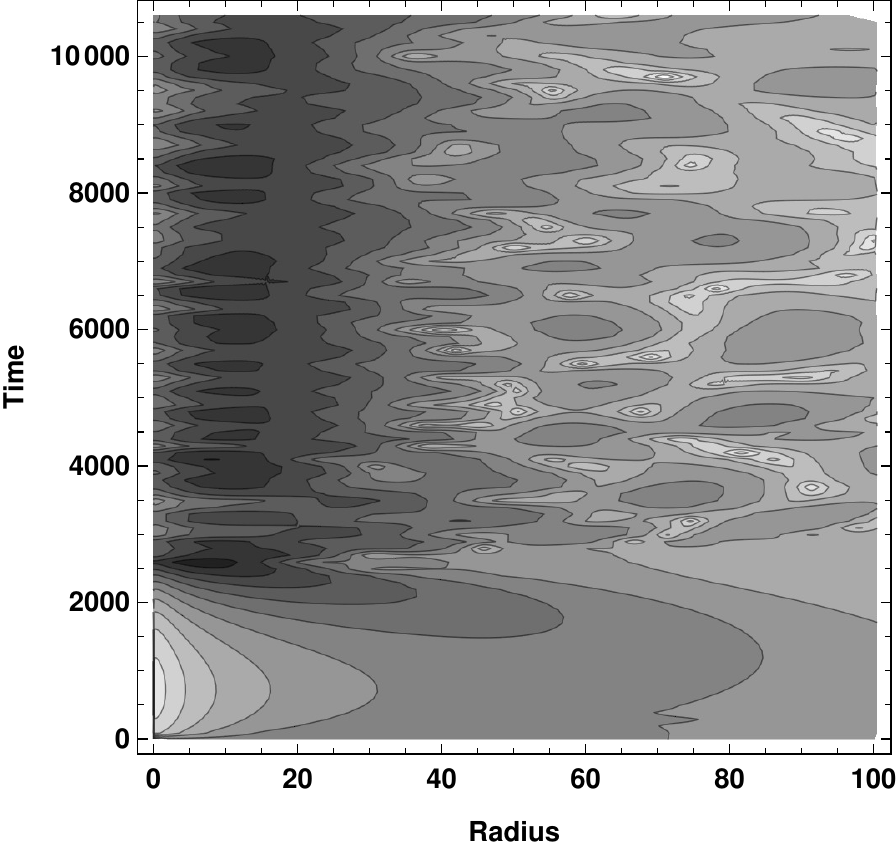} 
\caption{Spacetime diagram of the Proca energy density $\rho_e$ for model 3. This is a logscale plot and the darker areas represent zones where $\rho_e$ is higher.}
\label{fig:Density}
\end{minipage}
\end{figure}

Examining the magenta line in Fig.~\ref{fig:Energy} one observes some significant oscillations in the value of $E_{10}$. The existence of strong radial oscillations of the Proca energy density $\rho_e$ near the origin is confirmed in Fig.~\ref{fig:Density} where we display a spacetime diagram of $\rho_e$. This shows that $\rho_e$, which at $t=0$ is peaked around $r=0$, initially decreases near the origin, a signature that the Proca field is drifting away from the origin; but around $t\simeq 1200$ the expansion stops and the field starts to collapse. From $t\simeq 2000$ onwards, we see that the density profile is confined within the volume delimited by $r=30$,  and it undergoes radial oscillations. These oscillations dissipate part of the Proca energy to infinity in a similar process to that observed  in the scalar case~\cite{Seidel:1993zk}, which we therefore also dub ``gravitational cooling".  Further inspection of Fig.~\ref{fig:Density2}, in which we focus on the late-time evolution of the radial profile of the energy density of the Proca field, shows how the field oscillates radially and loses part of the energy in the process, as the height of the peak gradually decreases.

%
%
%\begin{figure}
%\begin{minipage}{1\linewidth}
%\includegraphics[width=1.0\textwidth]{img/At-eps-converted-to.pdf} 
%\includegraphics[width=1.0\textwidth]{img/Ar-eps-converted-to.pdf} 
%\caption{Evolution of the radial profile of $A_t$ (\textit{top panel}) and $A_r$(\textit{bottom panel}). In the inset we focus near the node of $A_t$.}
%\label{fig:At-Ar}
%\end{minipage}
%\end{figure}
%

%

\begin{figure}
%\begin{minipage}{1\linewidth}
\includegraphics[width=0.32\textheight]{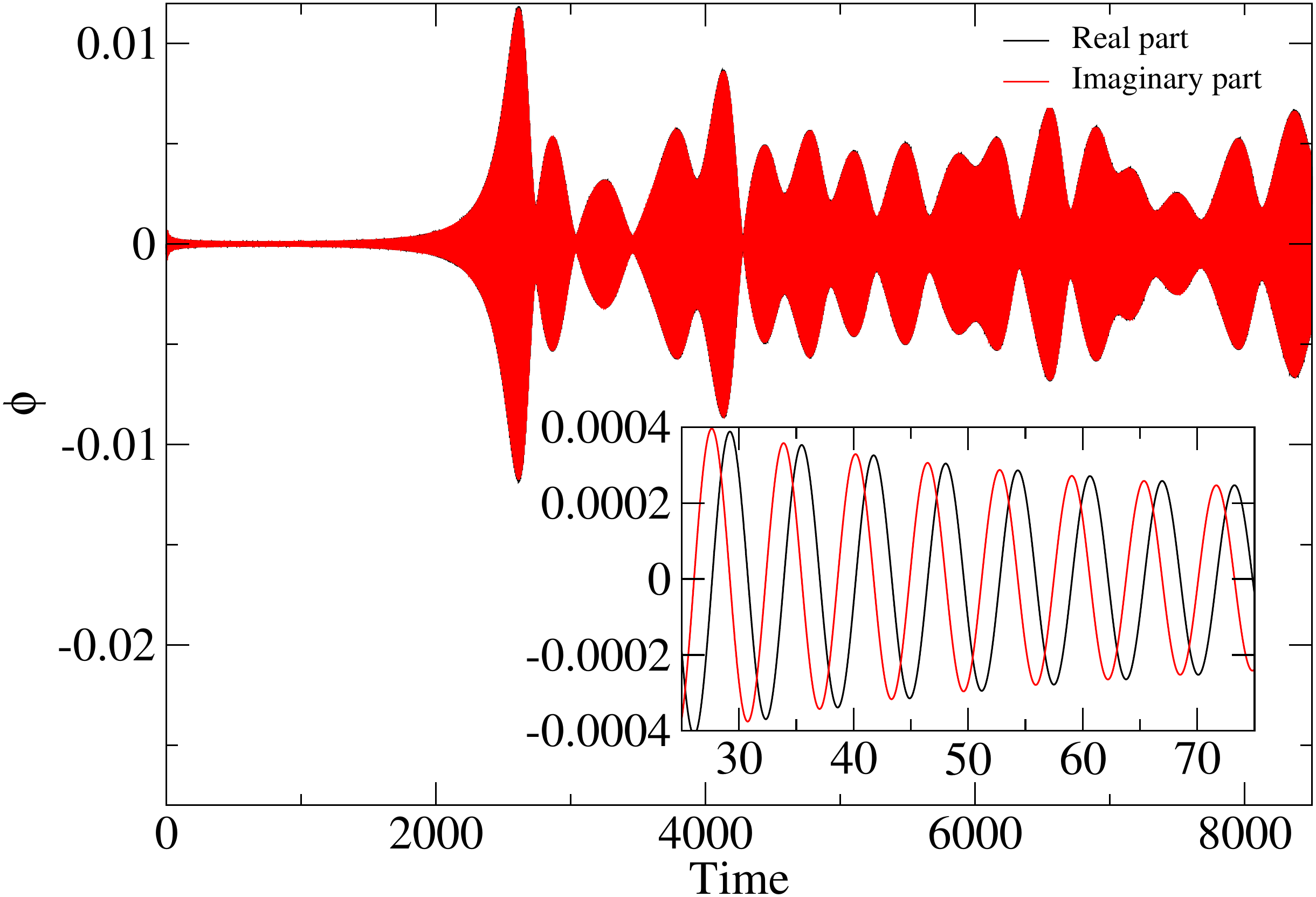}
\includegraphics[width=0.32\textheight]{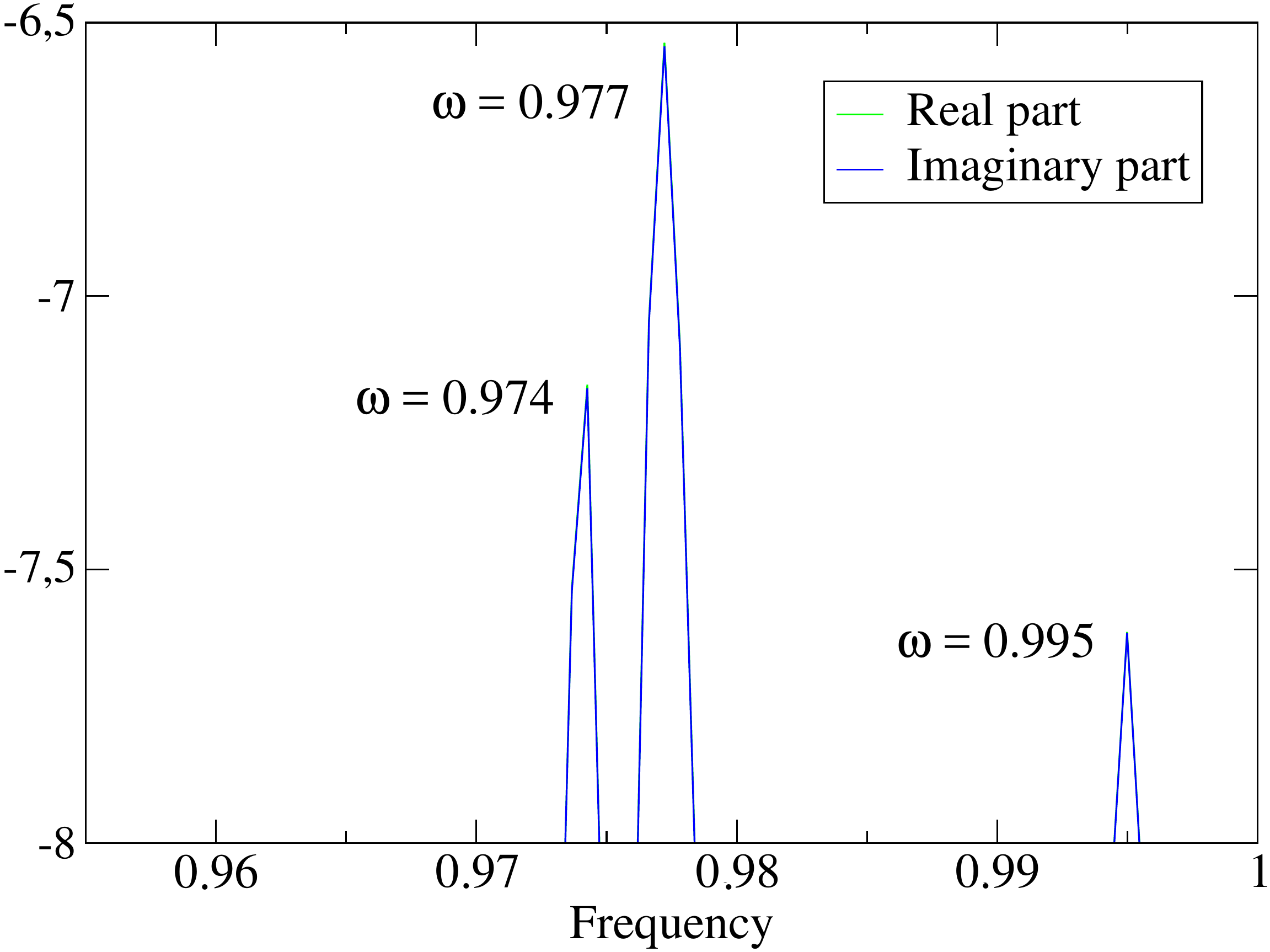} 
\caption{\textit{Top panel}: Time evolution of the real and imaginary parts of the scalar potential $\Phi$ at $r=5$ for model 3. The inset shows a shorter time window $t \in [25,75]$, where the ${\pi}/{2}$ phase difference becomes clear. \textit{Bottom panel}: Corresponding Fourier transform of the two time series of $\Phi$ in the top panel. The peaks for the real and imaginary parts completely overlap, so that only the blue line appears. The units in the vertical axis are arbitrary.}
\label{fig:phi}
%\end{minipage}
\end{figure}

So far we have been interested in studying components of the Proca field and its energy density, to establish the formation of a compact object. We can now further establish the final object is a Proca star by observing if the phase difference between the real and imaginary parts remains $\pi/2$ during the evolution, as requested for a Proca star.  Boson, Proca and oscillatons stars are defined in their ground state by a frequency at which the field
oscillates with time. A time series of the amplitude of the real and imaginary parts of the field at some
extraction radius can be built. The frequency can be then obtained by performing a Fourier analysis of the
series. In the case of bosonic stars, there is only one frequency. In our work, we studied the formation of a
Proca star in a highly dynamical situation; the compact object forming is perturbed. Nevertheless, we can
build the time series of the amplitude (top panel of Fig.~\ref{fig:phi}) of both the real and imaginary parts of the scalar potential $\Phi$ at a fixed radius $r=5$. In the inset we focus on a short time window and we confirm that the phase difference of ${\pi}/{2}$ which was set as an initial condition for model 3, remains the same value during the evolution. In the bottom panel of Fig.~\ref{fig:phi} we performed the Fourier transform to determine the oscillating frequency(-ies). A
beating pattern appears in the time evolution of the amplitude due to the three different frequencies observed for model 3. (All model frequencies are reported in Table~\ref{table:models}.) In this example, the FFT of the real and imaginary parts of the scalar potential yield the same exact frequencies, as expected. 

The time invariance of the initial phase difference still holds when considering more general phase differences $\delta$. In Fig.~\ref{fig:phase} we plot the four complex cases that we have studied (models 1 to 4). The phase difference during the evolution is computed by finding the times, from the data, at which the real and imaginary parts of $\Phi$ vanish, thus yielding the difference. Fig.~\ref{fig:phase} establishes that the phase differences are unchanged during the evolution. The small oscillations seen in the figure for the models with a phase offset $\delta={\pi}/{2}$ and $\delta={\pi}/{3}$  are due to numerical innacuracies induced by finding the zeros. In the cases with $\delta=0$ and $\delta=\pi$  we see a perfect horizontal line, because the profiles of the real and imaginary part coincide and the errors in finding the zeros precisely cancel each other. Therefore, for each initial $\delta$ we find an apparently stable solution which oscillates periodically with the same preserved initial phase difference. Our results are in agreement with those of \cite{Hawley:2002zn} for the scalar case, where  the existence of a continuous family of periodic soliton-like solutions for the Einstein-Klein-Gordon system (parametrized by the phase difference $\delta$) was conjectured. However, except for the solution with a $\pi/2$ phase difference one does not expect such soliton-like solutions to  be absolutely stable. Rather they are akin to oscillatons and slowly decay, albeit in extremely long timescales~\cite{Page:2003rd}. 

\begin{figure}[t]
\begin{minipage}{1\linewidth}
\includegraphics[width=0.9\textwidth]{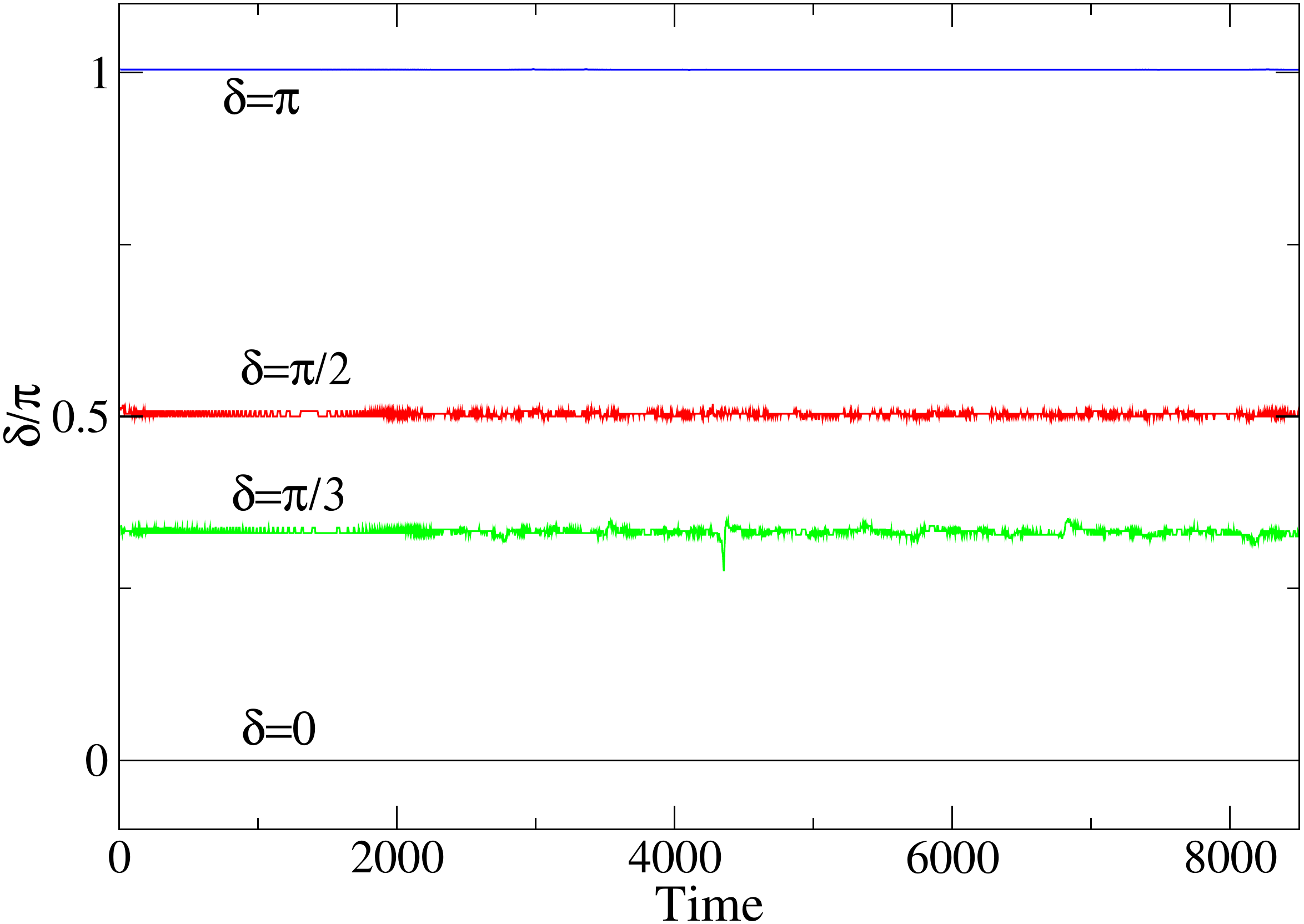} 
\caption{Phase difference $\delta$ between real and imaginary part of the scalar potential $\Phi$ calculated at radius $r=5$ for the models 1,2,3,4, corresponding, respectively, to an initial phased difference $\delta=0,{\pi}/{3},{\pi}/{2},\pi$.}
\label{fig:phase}
\end{minipage}
\end{figure}

\begin{figure}
\begin{minipage}{1\linewidth}
\includegraphics[width=0.9\textwidth]{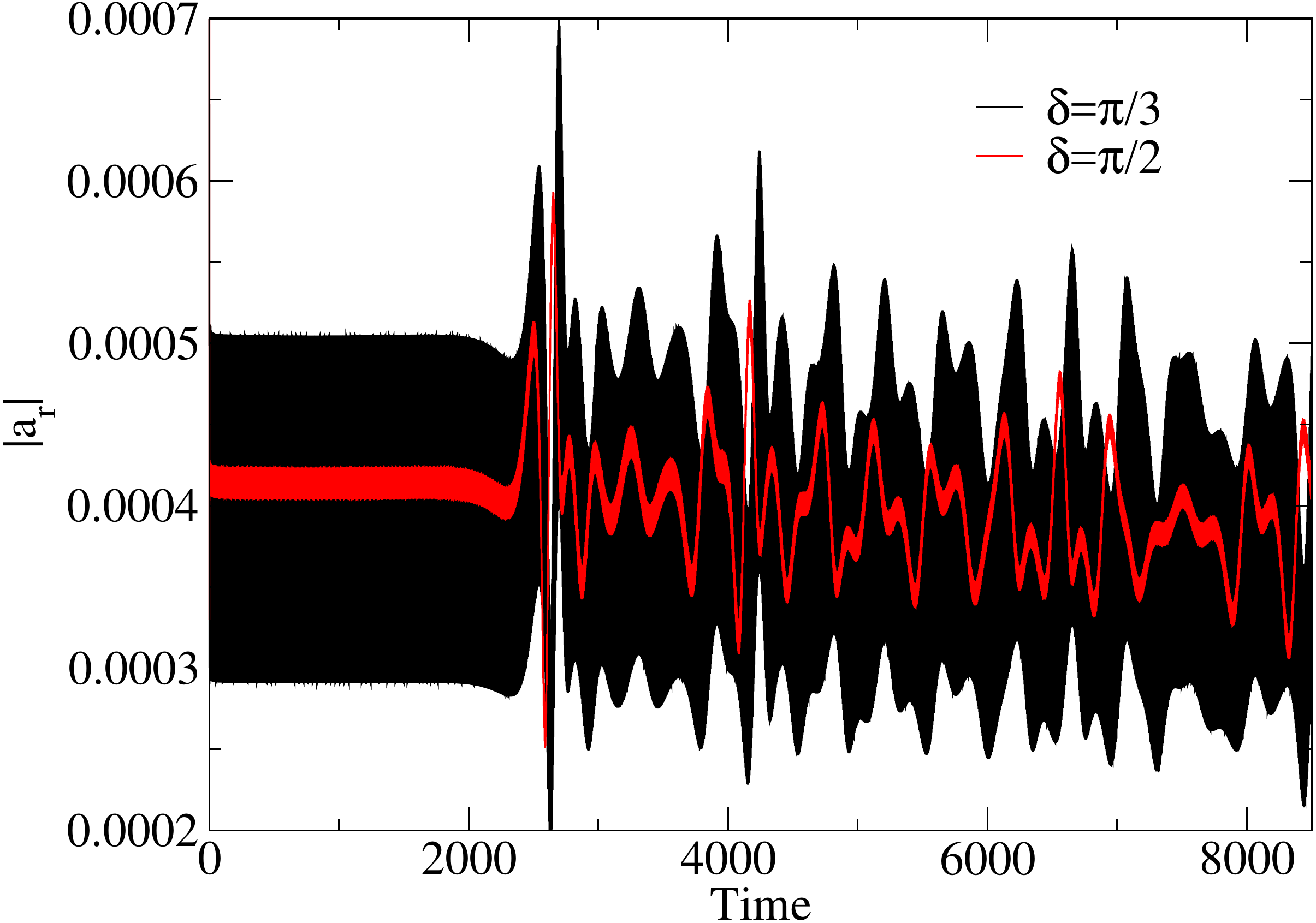} 
\caption{Time evolution of the norm of the vector potential amplitude, $|a_r|$, at $r=5$ for an initial phase difference of ${\pi}/{2}$ (model 3) and  ${\pi}/{3}$ (model 2). }
\label{fig:amplitude}
\end{minipage}
\end{figure}

Now we turn our attention to the way evolutions change if a different phase difference is chosen. In Fig.~\ref{fig:amplitude} we plot the norm of the amplitude of the vector potential $a_r$ evaluated at radius $r=5$ and defined by
\begin{equation}
|a_r|=\sqrt{\operatorname{Re}(a_r)^2+\operatorname{Im}(a_r)^2}\,,
\end{equation}
for models 2 and 3. For the case with $\delta={\pi}/{2}$ (model 3)  the amplitude  oscillations are small at early times ($t \in [0,2000]$). Only when the compact object starts to form, around $t\sim 2000$, more significant oscillations start to occur. For the case with $\delta={\pi}/{3}$ (model 2), the amplitude of the initial oscillations is significantly larger than for the previous case. Therefore, the $\pi/2$ phase difference appears to suppress the dynamics, as expected.

\begin{figure}
\begin{minipage}{1\linewidth}
\includegraphics[width=0.9\textwidth]{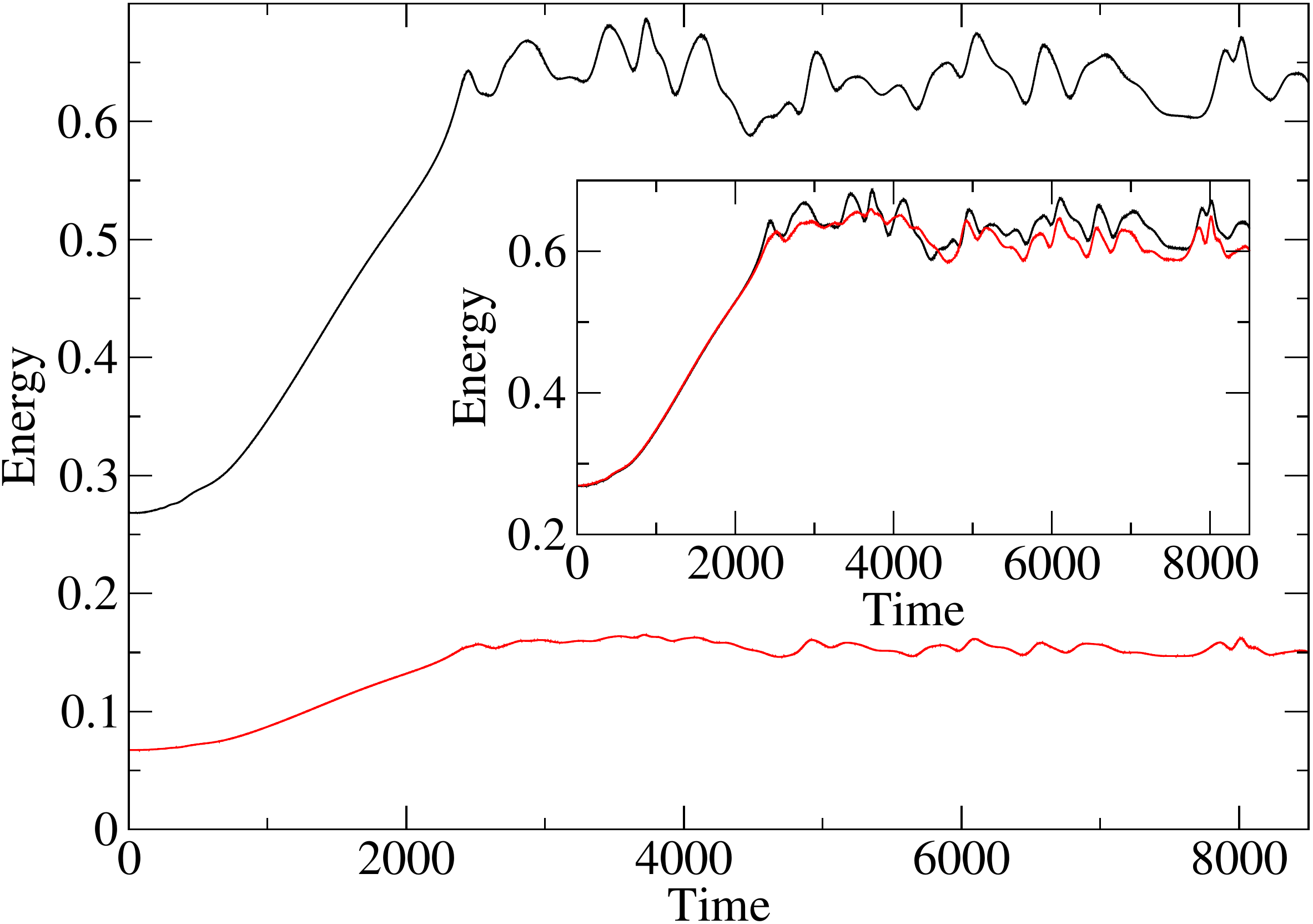} 
\caption{Time evolution of the energy contribution from the real part of the Proca field (black line) and imaginary part (red line), for model 6. In the inset we multiply the imaginary part by a factor 4 to compare the two curves.}
\label{fig:1}
\end{minipage}
\end{figure}

Furthermore, using models 6 and 7, we can study if the amplitude ratio $A_2/A_1$ is conserved during the dynamical formation of a Proca star. In Fig.~\ref{fig:1} we plot  the time evolution of the energy for model 6 inside a sphere of $r=100$. Each line of the main figure indicates the contribution given by either the real part or the imaginary part of the field. In the inset we multiply the imaginary part by a factor 4 to compare the two curves and we observe that they match well, within numerical accuracy, thus implying that the amplitude ratio is conserved. A similar behaviour is also found for model 7.

\begin{figure}
\begin{minipage}{1\linewidth}
\includegraphics[width=0.9\textwidth]{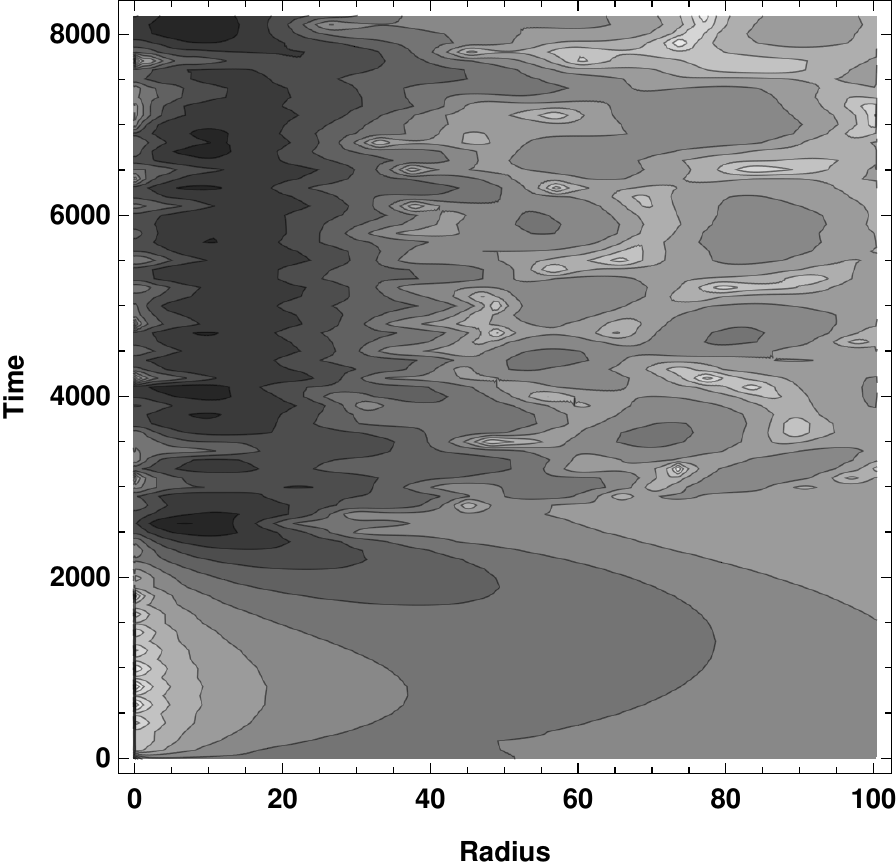} 
\caption{Spacetime diagram of the Proca energy density $\rho_e$ for the real oscillaton model 5. This is a logscale plot and the darker areas represent zones where $\rho_e$ is higher.}
\label{fig:Density_real}
\end{minipage}
\end{figure}

Finally, we use model 5 to study the real oscillaton case. Our results show that the behaviour of this model resembles that of the complex cases described above. As an illustration, we exhibit in Fig.~\ref{fig:Density_real} the spacetime diagram of  $\rho_e$ for this model, similarly to what we have done in Fig.~\ref{fig:Density} for the complex case. A clear parallelism can be seen between the two cases, and, in particular, the gravitational cooling mechanism is at work also in the real Proca case.

%%%%%%%%%%%%%%%%%%%%%%%%%%%%%%%%%%%%%%%%%%%%%%%%%%%
\section{Conclusions} \label{sec:conclusions}
%%%%%%%%%%%%%%%%%%%%%%%%%%%%%%%%%%%%%%%%%%%%%%%%%%%

In this work we have studied the dynamical formation of spherical Proca stars, as well as of quasi-stationary vector solitonic objects, by performing fully non-linear numerical relativity simulations, within the Einstein-(complex)Proca theory. For this purpose, we succeeded in obtaining appropriate initial data that solves all three constraints of the system, but that still leaves enough freedom to vary the initial cloud parameters as to probe the dependence of the process on the phase difference between the real and imaginary parts of the Proca field and on the amplitude ratio between these two components. Whereas truly stationary solitonic objects - Proca stars~\cite{Brito:2015} - only exist for a phase difference of $\pi/2$ and equal amplitudes, quasi-stationary vector solitons - analogous to scalar oscillating soliton stars~\cite{Seidel:1991zh} - appear to exists for more general phase differences and amplitude ratios. Indeed, our simulations suggest that in these cases, the phase difference and amplitude ratio are conserved throughout the simulations and, nonetheless, a compact object tends to form. Thus, we conjecture the existence of a continuous family of quasi-stationary vector solitons, similar to that discussed in~\cite{Hawley:2002zn} for the scalar case. 

As an avenue for future research, the formation of \textit{rotating} boson or Proca stars from the gravitational collapse of a (rotating) cloud of scalar/vector matter, remains an outstanding open question.

%%%%%%%%%%%%%%%%%%%%%%%%%%%%%%%%%%%%%%%%%%%%%%%
\section*{Acknowledgements}
%%%%%%%%%%%%%%%%%%%%%%%%%%%%%%%%%%%%%%%%%%%%%%%

F.D.G. thanks Pierre Pizzochero for his helpful suggestions. We would also like to thank E. Radu for discussions. C.H. acknowledges funding from the FCT-IF programme.  This work was supported by the Spanish MINECO (AYA2015-66899-C2-1-P), by the Generalitat Valenciana (PROMETEOII-2014-069, ACIF/2015/216), and by the European  Union's  Horizon  2020  research  and  innovation  programme  under  the H2020-MSCA-RISE-2015 Grant No.   StronGrHEP-690904, the H2020-MSCA-RISE-2017 Grant No. FunFiCO-777740  and  by  the  CIDMA  project UID/MAT/04106/2013. The authors also  acknowledge networking support by the COST Action GWverse CA16104. All computations have been performed at the Servei d'Inform\`atica de la Universitat de Val\`encia.

%%%%%%%%%%%%%%%%%%%%%%
%%%   REFERENCES   %%%
%%%%%%%%%%%%%%%%%%%%%%

\bibliography{biblio}

\end{document}